\documentclass[aps,pra,twocolumn,superscriptaddress]{revtex4}
\usepackage{amsmath}
\usepackage{amsbsy,latexsym}
\usepackage{amsfonts}
\usepackage{amssymb}
\usepackage[mathscr]{eucal}
\usepackage{graphicx}
\newcommand{\rbf}{\mathbf{r}}
\newcommand{\Rbf}{\mathbf{R}}
\newcommand{\Vbf}{\mathbf{V}}

\newcommand{\tr}{{\rm tr}\,}

\newcommand{\NN}{{\mathscr N}}

\begin{document}
\voffset 1.5cm
\title{Collective vs local measurements in qubit mixed state estimation  }

\author{E.~Bagan}
\author{M.~Baig}
\author{R.~Mu{\~n}oz-Tapia}
\affiliation{Grup de F{\'\i}sica Te{\`o}rica \& IFAE, Facultat de Ci{\`e}ncies,
Edifici Cn, Universitat Aut{\`o}noma de Barcelona, 08193 Bellaterra
(Barcelona), Spain}
\author{A. Rodriguez}
\affiliation{Grup de F{\'\i}sica Te{\`o}rica \& IFAE, Facultat de Ci{\`e}ncies,
Edifici Cn, Universitat Aut{\`o}noma de Barcelona, 08193 Bellaterra
(Barcelona), Spain}
 \affiliation{Area de F{\'\i}sica Te{\'o}rica, Facultad
de Ciencias, Universidad de Salamanca, 37008 Salamanca,  Spain}
%\date{06/05/2002}

\begin{abstract}
We discuss the problem of estimating a qubit mixed state. We give
the optimal estimation that can be inferred from any given set of
measurements. For collective measurements and for a large number,
$N$, of copies, we show that the error in the estimation goes as
$1/N$. For  local measurements, we focus on the simpler case of
states lying on the equatorial plane of the Bloch sphere. We show
that the error using plain tomography goes as $1/N^{1/4}$, while
our approach leads to an error proportional to $1/N^{3/4}$.

\end{abstract}

\maketitle
\section{Introduction}
Knowing the state of a system is of para\-mount importance in
quantum information. Quantum measurements provide only a partial
knowledge of a state. Such a state, can only be reasonably
reconstructed if a large number, $N$, of identically prepared
copies of the system is available. Since the seminal work of
Holevo~\cite{holevo} there has been a lot of research on this
subject. Most of the quantitative analysis have mainly focused on
pure states, for which the optimal
strategies~\cite{mp,jones,pure-states,us-product,us-local} have
been identified.  They give the ultimate limits that can be
achieved in state reconstruction. However, they involve collective
measurements that, although very interesting from the theoretical
point of view, are very resource consuming and very difficult to
implement in a lab.
%Many important features have also been
%identified, one of the most important being that the quantum
%correlations involved in the collective measurements are more
%efficient than the classical correlations of local
%measurements\cite{pw,mp}.

In the real world pure states are very scarce and so mixed state
estimation is not just an academic issue.
%Quite on the contrary, it is experimentally relevant.
For instance, it is important to estimate the purity of a state,
since this parameter often determines its utility  to perform
quantum information tasks. In quantum tomography, a
\textit{quorum} of local observables is measured on a (large)
number of copies of a state $\rho$. From the relative frequencies
of the outcomes, one then obtains an approximation or guess
$\rho_g$ to the signal state $\rho$. However, the statistical
deviations often yield unphysical states, e.g. $\tr{\rho_g}> 1$.
In this case, one can either discard the results or use a maximum
likelihood data analysis \cite{likelihood}. Using this analysis
one infers the physical state that provides the closest
theoretical probabilities to the observed frequencies. Many
variants of these techniques can be found in the
literature~\cite{tomography,likelihood}, but there is a notorious
lack of quantitative results (see though \cite{vidal,density}).

The large numbers law ensures that with an infinite number of
copies ($N \to \infty$) and infinite measurements the state could
be exactly reconstructed by any sensible method. In practice,
however, one has access only to limited resources and the  crucial
issue is to quantify the quality of the reconstruction procedure.
This is the question we address here. We focus on  qubit states
and  use the fidelity as  figure of merit. We obtain the best
estimate for any given measurement and  compute the analytical
expressions of the average fidelity for both  collective and local
(von Neumann) measurements in the asymptotic limit (large $N$).

%We have recently shown that for \emph{pure} states estimation
%local measuring strategies~\cite{us-local} are very competitive
%with respect to (complicated) collective measurements. However
%this not the case here. In particular, a naive tomographic
%reconstruction of the estate yields an error that is at least a
%fourth power worse in the number copies than the optimal
%collective strategy. We will show how this result can be
%ameliorated, but contrary to what happens with pure states, fixed
%local measuring strategies still yield an error qualitatively
%larger than collective measurements.

%The paper is organised as follows. In section II we recall some
%basic formulae of the fidelity and obtain the uniform probability
%density for qubits. In section III we give a compact expression of
%the average fidelity. In section IV we present/illustrate our
%results for a fixed measuring strategy and for a state
%reconstructed from the tomographic, the maximum likelihood and
%optimal guess. Section V contains brief conclusions and outlook
%for further work.

\section{Bures Metric}

The estimation procedure goes as follows.  After we have measured
on the $N$ copies of the system, some result is obtained, which we
symbolically denote by $x$. Note that $x$  stands for both a
single outcome of a collective measurement or a list of $N$
outcomes (one for each individual measurement) in local schemes.
Based on $x$, an estimate for $\rho$ can be guessed, $\rho_g(x)$.
The fidelity is defined as~\cite{bures}
\begin{equation}\label{f}
 f=\tr\sqrt{{\rho_g^{1/2}(x)}\; \rho \; {\rho_g^{1/2}(x)}}.
\end{equation}
It determines the maximum distinguishability between $\rho$ and
$\rho_g(x)$ that can be achieved by any measurement \cite{fuchs}.
For qubits, Eq.~1 reads
\begin{equation}\label{f-qubits}
 f(\vec{r},\vec{R}(x))=\frac{1+\vec{r}\cdot
 \vec{R}(x)+\sqrt{1-r^2}\sqrt{1-R(x)^2}}{2}.
\end{equation}
Here $r=|\vec{r}|$ and $R=|\vec{R}|$, where $\vec{r}$  and
$\vec{R}(x)$ are the Bloch vectors of the states $\rho$ and
$\rho_g(x)$ respectively [$\rho=(1+ \vec{r}\cdot \vec{\sigma})/2$;
$\vec{\sigma}=(\sigma_x,\sigma_y,\sigma_z)$ are the standard Pauli
matrices].

The fidelity can be viewed as a \lq\lq distance" between two
density matrices. The corresponding metric is usually known as
Bures metric. From the infinitesimal \lq\lq distance"
$f(\vec{r},\vec{r}+d\vec{r})$ it is easy to obtain the volume
element (normalized to unity: $\int \, d\rho =1$)
\begin{equation}\label{drho3}
 d\rho=\frac{2}{\pi}\frac{r^2 dr}{\sqrt{1-r^2}}d n, \quad
 dn=\frac{\sin \theta d\theta d\phi}{4 \pi},
\end{equation}
 where $dn $
is the invariant measure in the two-sphere. Eq.~\ref{drho3} is the
natural \emph{uniform} probability distribution function, or the a
priori probability distribution for a \emph{completely unknown}
qubit state $\rho$. For our discussion below, we will also need
$d\rho$ when the density matrices are known to lie in a great
circle of the Bloch sphere. It reads
\begin{equation}\label{drho2}
 d\rho=\frac{1}{2 \pi}\frac{r dr}{\sqrt{1-r^2}}d \theta.
\end{equation}
Although in Section~\ref{results} we use \eqref{drho3} and
\eqref{drho2}, our main results in the following section are
independent of  any particular choice of the a priori
distribution.

\section{Fidelity and Optimal Guess}
The average fidelity, hereafter fidelity in short, is the mean
value of (\ref{f}) over the a priori distribution  and over all
possible outcomes $x$,
\begin{equation}\label{fidelity}
     F=\sum_x\int d\rho \, f(\vec{r},\vec{R}(x)) p(x|\vec{r}),
\end{equation}
where $p(x|\vec{r})$ is the conditional probability of obtaining
outcome $x$ if the signal state has Bloch vector $\vec{r}$. These
probabilities are determined by the expectation values of positive
operators $O(x)$, such that $\sum_x O(x)=\mathbb{I}$, i.e.,
 $p(x|\vec{r})=\tr [O(x) \rho]$. Our aim is to maximize
 \eqref{fidelity}.

 For a given measurement scheme, $\{O(x)\}$, there always exists
 an  optimal guess, as we now show.
 We first  introduce the four
dimensional Euclidean vector
\begin{equation}\label{rmu}
    \mathbf{r}=(\sqrt{1-r^2},\vec{r}).
\end{equation}
Note that $\rbf\cdot
\rbf'=\sqrt{1-r^2}\sqrt{1-r'^2}+\vec{r}\cdot\vec{r}\,'$  and
$|\rbf|=\sqrt{\rbf\cdot\rbf}=1$.  With this, the  average fidelity
reads
\begin{equation}\label{fidelity-2}
     F=\sum_x\int d\rho \, \frac{1+\rbf\cdot \Rbf(x)}{2}
     p(x|\vec{r}).
\end{equation}
A straightforward use of the Schwarz inequality gives an upper
bound of $F$ that is saturated with the choice
\begin{equation}
  \Rbf(x)= \frac{\Vbf(x)}{|\Vbf(x)|};\quad \Vbf(x)=\int d\rho\, \rbf \, p(x|\vec{r}).
    \label{vector-1}
\end{equation}
Using \eqref{vector-1}, the maximum fidelity is
\begin{equation}
    F=\frac{1}{2}\left(1+\sum_{x}|\Vbf(x)|\right).
    \label{fidelity-general}
\end{equation}
Since  the guess \eqref{vector-1} satisfies $|\Rbf(x)|=1$ and its
first component is non-negative, it \emph{always} gives a physical
state. In fact, for any set of measurements and any a priori
distribution, \eqref{vector-1} is the best state that can be
inferred and \eqref{fidelity-general} is the maximum fidelity.

As the number of copies of the system becomes asymptotically
large, any reasonable  estimation scheme leads to a perfect
reconstruction of the state, i.e, $F\to 1$. For a large, but
finite $N$, the relevant issue is knowing the rate at which  the
perfect estimation limit is attained.  For pure states, it is well
known that the best collective strategy yields $F\sim 1-1/N$
($F=1-1/(4 N)$ for states on the equator of the Bloch sphere)
\cite{mp}. It has also been  shown recently that this asymptotic
limit can be achieved with local measurements~\cite{us-local}. For
mixed states much less is known. Most of the pure state results
can not be extrapolated to the mixed case, and some others may
look counterintuitive at first sight. Although the space of mixed
states seems to be larger, they are less distinguishable than pure
states. The fidelity (\ref{f-qubits}) has a minimum value
$(1-r)/2$ which is never zero but for pure states ($r=1$). Thus
the average fidelity could, in principle, be larger than that of
pure states alone. Note that  any estimated mixed state $\rho_g$
has some overlap with the signal state $\rho$. To be more
concrete, imagine one does random guessing, without performing any
measurement at all, i.e., $p(x|\vec{r})$ is uniform. Then, using
Bures volume element~\eqref{drho3}, the average fidelity is
$F_{\rm rand}=1/2+8/(9 \pi^2)$, which is larger than the random
value ($F=1/2$) for pure states.
% reads in a self explanatory notation
%\begin{equation}\label{random}
% F_{\rm rand}= \int d \rho_R
%   \int d \rho_r  \, f(\vec{r},\vec{R})= \frac{1}{2}+\frac{8}{9 \pi^2},
%\end{equation}

\section{Results}\label{results}
\subsection{Collective measurements}
 As a first application of the results of the previous section,  let
us obtain the asymptotic behavior of the fidelity with the optimal
collective measurement scheme. The main results are contained in
\cite{vidal}, where an optimal (and minimal) generalised
measurement was obtained for qubit density matrices and generic
isotropic probability distributions. However no definite form for
this distribution was assumed and no explicit results were
obtained. Our approach enables us  to cast the expressions in
\cite{vidal} in a more transparent way as well as to simplify some
of the derivations there.
%For a collection of $N$ identically
%prepared states $\rho$,
The optimal measurement is represented by a set of positive
operators and conditional probabilities that  can be labelled by
two indices $x=(k,\vec{m})$. The  discrete index $k$ labels the
representations of the symmetric space spanned by $\{\rho^{\otimes
N}\}$ onto which the positive operators of the measurement
project, whereas  the unit vector $\vec{m}$ labels a continuous
set of outcomes in the two-sphere~\cite{comentari}. We have
\begin{equation}\label{collective probablilites}
    p(k,\vec{m}|\vec{r})=
    %&\sum_i\tr[O_{k,i}^{(N)} \rho^{\otimes N}]\nonumber  &=&\\
    c_k\left(\frac{1-r^2}{4}\right)^{N/2-k}
     \left(\frac{1+ \vec{r}\cdot
    \vec{m}}{2}\right)^{2k},
\end{equation}
where
\begin{equation}\label{ck}
c_k= \left(
      \begin{array}{c} N\\
                      N/2+k
      \end{array}
\right) \frac{(2k+1)^{2}}{N/2+k+1}.
\end{equation}
From (\ref{vector-1}) and (\ref{fidelity-general}) we obtain
\begin{equation}\label{fidelity-vidal}
    F=\frac{1}{2}+\frac{1}{2}\sum_k \int d m\,
    |\Vbf(k,\vec{m})|.
\end{equation}
The sum in  \eqref{fidelity-vidal} runs from $k=0$ ($k=1/2)$ for
$N$ even (odd) to $k=N/2$. Taking advantage of the rotational
invariance, the integrals in (\ref{fidelity-vidal}) can be
evaluated exactly. The computation of the  asymptotic limit is
rather lengthy and will not be reproduced here~\cite{prep}. The
final result is
\begin{equation}\label{fasymptotic}
    F=1-\left(\frac{3}{4} +\frac{4}{3\pi}\right)\frac{1}{N}+\cdots\ .
\end{equation}
Note that this fidelity is only slightly  worse than that of  pure
states: $3/4 + 4/(3\pi)=1.17\gtrsim 1$.

%The specific value of the coefficient of the $1/N$ term  may vary
%for other uniform isotropic distributions, i.e. of the form
%$d\rho=w(r)dr dn$ with $w(r)$ a normalized positive function,  but
%the linear behavior in  $1/N$  of the asymptotic limit is expected
%to remain.

 The expression
(\ref{fasymptotic}) also gives us important information about the
optimal fidelity when the \textit{a priori} probability
distribution corresponds to states known to lie in the equator
plane of the Bloch sphere (\ref{drho2}). Since in this situation
we have more information about the states, the fidelity cannot be
worse than (\ref{fasymptotic}), i.e. the error, defined as
$E=1-F$, must satisfy $E\leq \xi/N $, where $\xi$ is a constant.
%(we have numerical hints that $\beta=1$, but we have
%not proved it rigorously).

\subsection{Local measurements}
Let us now tackle the problem of reconstructing a qubit state
$\rho$ from local measurements.  The analytical expressions turn
out te be rather difficult to obtain and for simplicity only the
case of  states that are known to lie on the equator plane of the
Bloch sphere~\eqref{drho2} will be considered in this note. This
is a non trivial case that can be relevant for quantum optics
(e.g., for polarization states of photons). We will only sketch
our main results. The techniques we have used are essentially
contained in~\cite{us-product,us-local}. Full details of the
calculations will be presented elsewhere~\cite{prep}.

Consider $N=2\NN$ copies of the state $\rho$. Quantum state
tomography tells us that von Neumann measurements along two fixed
orthogonal directions, $x$ and $y$, are sufficient to reconstruct
the state. After the measurements, we obtain a set of outcomes
$+1$ and $-1$ with relative frequencies $\alpha_i$ and
$1-\alpha_i$, respectively ($i=x,y$). This occurs with probability
\begin{equation}\label{probablity}
  p(\vec{\alpha}|\vec{r})=  \prod_{i}\left(\begin{array}{c}
                              \NN\\
                              \NN \alpha_i
                       \end{array}
                  \right)\left(\frac{1+r_i}{2}\right)^{\NN\alpha_i}\!\!
                  \left(\frac{1-r_i}{2}\right)^{\NN(1-\alpha_i)}\kern-2em.
\end{equation}
 In quantum tomography the guess is given by
\begin{eqnarray}\label{guess-tomography}
    \Rbf_{\rm T}(\vec{\alpha})=(\sqrt{1-R^2},
   R\cos \gamma,R\sin \gamma),\\
R\cos\gamma=2\alpha_x-1,\ \  \ \ \ R\sin\gamma=2\alpha_y-1
.\nonumber
\end{eqnarray}
In many instances, however, the statistical fluctuations produce
an unphysical guess (the square root term becomes imaginary). If
one discards these cases, the asymptotic behavior of the fidelity
can be shown to be: $F=1-\xi_{\rm T}/N^{1/4}+\cdots$,
%\begin{equation}\label{error-tomography}
%    E=1-F=\frac{\alpha}{N^{1/4}}+\cdots
%\end{equation}
where $\xi_{\rm T}$ is a constant. Although plain tomography
yields a perfect reconstruction of the state in the asymptotic
limit, it is  much worse than the optimal collective scheme (note
the power 1/4 of $N$ as compared to the power 1 in
Eq.~\ref{fasymptotic}). One may suspect that the culprit of this
behavior is the number of copies discarded, but we now show that
it is not entirely so.

Within the maximum likelihood framework~\cite{likelihood} all
available data is used. If $R\leq1$, the guess is the tomographic
one: $\Rbf_{\rm ML}=\Rbf_{\rm T}$ (see
Eq.~\ref{guess-tomography}), and
\begin{equation}\label{guess-ml}
     \Rbf_{\rm ML} = \left(0,\cos(\Phi),\sin(\Phi)\right)
\end{equation}
if $R>1$, where $\Phi$ is the solution of the equation
$\cos(2\Phi)=R\cos(\gamma+\Phi)$. In the asymptotic limit one can
expand this equation as a power series in  $(R-1)$,
$\Phi=\gamma-(R-1)\cot 2\gamma+\cdots$. In fact only the first
term is necessary for our calculation.
%The maximum likelihood strategy to
%zeroth order consists in taking the central limit guess whenever
%it lies within the unit circle and its natural projection onto the
%circle if it lies outside.
After some effort one gets
\begin{equation}\label{fidelity-ml}
F=1-\frac{\xi_{\rm ML}}{N^{3/4}}+\cdots;\quad \xi_{\rm
ML}=\frac{\Gamma(1/4)^3}{ 2^{5/4}9 \pi^2}\simeq 0.2256.
\end{equation}
Notice the  significant increase in the rate at which the fidelity
approaches unity as compared to plain tomography.

Finally, we have computed the fidelity for the optimal guess
\eqref{fidelity-general}. Here again, all available data is used
to produce a reconstruction of the state. In Fig.~1 we compare the
optimal guess and maximum likelihood methods for up to $N=20$
copies of a state. It is clear that the optimal guess strategy
always performs better. We have also obtained the asymptotic
limit. The fidelity reads in this case
\begin{equation}\label{fidelity-og}
    F=1-\frac{\xi_{\rm O}}{N^{3/4}}+\cdots; \quad \xi_{\rm O}\simeq
    0.1708,
\end{equation}
where $\xi_{\rm O}$ can be computed analytically (see appendix).
Notice that this fidelity approaches unity at a rate similar to
the maximum likelihood one~\eqref{fidelity-ml}, but the
coefficient of the first correction is lower ($\xi_{\rm
O}<\xi_{\rm ML}$), as it should.  The most important parameter is
the exponent of the $1/N$ term in (\ref{fidelity-og}). It shows
that, there is a gap in the quality of the reconstruction process
between fixed local measurements and optimal collective schemes
(recall Eq.~\ref{fasymptotic}). One may argue that we have not
exploited classical communication, i.e, we have not designed each
individual measurement according to the outcomes of the previous
ones. We have also explored this possibility numerically and
observed that the fidelity  is almost identical to that obtained
from the optimal guess with measurements along two fixed
orthogonal directions. Therefore, we are led to conjecture that an
error rate $E\sim 1/N^{3/4}$ is the lowest that can be achieved
using any local scheme.

 \begin{figure}
 \includegraphics[width=8cm]{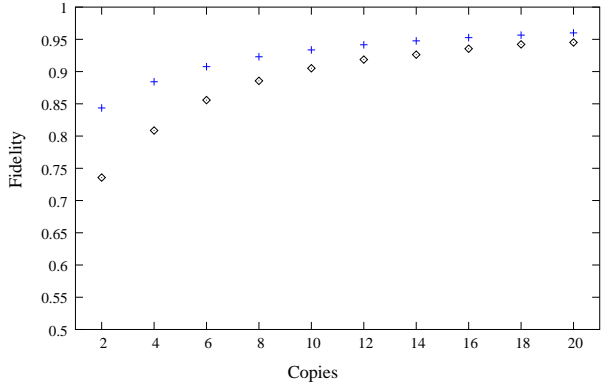}
 \caption {Average fidelities in terms of the number of copies.
 The diamonds (crosses) correspond to the maximum likelihood (optimal) guess.}
 \end{figure}

\section{Conclusions}
We have obtained the optimal reconstruction of a general qubit
state for any given set of measurements and  have illustrated our
results in some interesting cases. We have computed the asymptotic
expression of the fidelity for the optimal collective scheme. For
local measurements we have considered the simpler but important
case of states lying on the equator plane of the Bloch sphere. We
have shown that the performance of plain tomography  is very poor,
with an error that goes as $E\sim 1/N^{1/4}$ for large $N$. We
have shown that maximum likelihood  does provide a much better
estimation: $E\sim 1/N^{3/4}$. Using the same data, the optimal
guess analysis gives the best reconstruction of the signal state.
Despite this improvement, the asymptotic behavior of the fidelity
does not saturate the optimal collective bound. This is in
contrast to  pure state estimation, where local measurements can
perform optimally in the asymptotic regime. Although we have
mainly focussed on measurements along fixed orthogonal directions,
we have also analysed the most general local strategy, in which
one is entitled to change these directions after each individual
measurement. Our results strongly suggest that for mixed states
the asymptotic behavior of the optimal collective schemes cannot
be attained by any local strategy.
\section*{Acknowledgements}
We thank A. Acin, G.M.~D'Ariano and  C. Mac\-chia\-vello  for
useful conversations. We acknowledge financial support from MCyT
projects BFM2002-02588 and BFM2002-02609,  CIRIT project
SGR-00185, and QUPRODIS working group EEC contract IST-2001-38877.
RMT thanks the hospitality of the Benasque Center for Science. ARG
thanks the hospitality of the IFAE and specially of the GIQ.

\appendix*
\section{}
The $\xi_{\rm O}$ constant in (\ref{fidelity-og}) is the sum of
three terms that can be written as
\begin{eqnarray*}
    \xi_{\rm O}&=&\frac{\Gamma(1/4)^2}{48 \pi}(4
b_1+b_2 -\sqrt{2}b_3);\\
 b_1&=&\int_{0}^{\infty}dx
\left(\frac{e^x}{x^{3/4}{\rm
Im}[K_{1/4}(-x)]}+\frac{\sqrt{2}}{\sqrt{\pi}x^{1/4}}\right),\nonumber \\
b_2&=&\int_{0}^{\infty}dx \frac{e^x [{\rm erfc}(\sqrt{2 x})-4]{\rm
erfc}(\sqrt{2 x})}{x^{3/4}{\rm Im}[K_{1/4}(-x)]}\nonumber, \\
b_3&=&\int_{0}^{\infty}dx \frac{e^x {\rm erfc}^2(\sqrt{2
x})}{x^{3/4}K_{1/4}(x)},
\end{eqnarray*}
where  $K_\nu$ and ${\rm erfc}$ are the modified Bessel function
and complementary error function respectively \cite{abramowitz}.
The integrals can be evaluated numerically: $b_1=0.197241$,
$b_2=1.61451$ and $b_3=0.31400$. With these numbers we have
$\xi_{\rm O}\simeq 0.17083$.

\end{document}